# Yb–Yb correlations and crystal-field effects in the Kondo insulator YbB$_{12}$ and its solid solutions

P A Alekseev[1], J-M Mignot[2], K S Nemkovski[1], E V Nefeodova[1],

N Yu Shitsevalova[3], Yu B Paderno[3], R I Bewley[4], R S Eccleston[4],

E S Clementyev[1,5], V N Lazukov[1], I P Sadikov[1] and N N Tiden[1]

[1] Russian Research Centre "Kurchatov Institute", 123182 Moscow, Russian Federation
[2] Laboratoire Léon Brillouin (CEA-CNRS), CEA/Saclay, F-91191 Gif sur Yvette, France
[3] Institute for Problems of Material Science, NASU, 252142 Kiev, Ukraine
[4] ISIS, Rutherford Appleton Laboratory, Didcot, Oxon, OX110QX, UK
[5] Physics Department E21, Technical university of Munich, D-85748 Garching, Germany

**Abstract**
We have studied the effect of Lu substitution on the spin dynamics of the Kondo insulator YbB$_{12}$ to clarify the origin of the spin-gap response previously observed at low temperature in this material. Inelastic neutron spectra have been measured in Yb$_{1-x}$Lu$_x$B$_{12}$ compounds for four Lu concentrations $x = 0$, 0.25, 0.90 and 1.0. The data indicate that the disruption of coherence on the Yb sublattice primarily affects the narrow peak structure occurring near 15-20 meV in pure YbB$_{12}$, whereas the spin gap and the broad magnetic signal around 38 meV remain almost unaffected. It is inferred that the latter features reflect mainly local, single-site processes, and may be reminiscent of the inelastic magnetic response reported for mixed-valence intermetallic compounds. On the other hand, the lower component at 15 meV is most likely due to dynamic short-range magnetic correlations. The crystal-field splitting in YbB$_{12}$ estimated from the Er$^{3+}$ transitions measured in a Yb$_{0.9}$Er$_{0.1}$B$_{12}$ sample, has the same order of magnitude as other relevant energy scales of the system and is thus likely to play a role in the form of the magnetic spectral response.

**Short title** : *Yb–Yb correlations and crystal-field effects in YbB$_{12}$ and its solid solutions*

## 1. Introduction

A significant number of strongly correlated electron systems containing lanthanide ($Ce_3Bi_4Pt_3$ [1], $SmB_6$ [2–4], TmSe [5, 6], $YbB_{12}$ [7]) or transition (FeSi [8]) elements have been found to develop an unconventional insulating ground state when cooled down below a characteristic temperature of the order of 100 K. While various physical properties give strong evidence for the opening of a narrow gap (~10 meV) in the electronic density of states, the exact role played by many-body effects (electronic correlations) is still a matter of debate [9]. One of the basic questions [10] is whether the gap formation can be understood in terms of strongly renormalized quasiparticle band parameters or, on the contrary requires other mechanisms beyond the scope of simple band theory, such as localization or short-range magnetic correlations, to be taken into account. A variety of theoretical approaches have been elaborated along both lines, making it an urgent task for experimentalists to determine more precisely the character of this insulating state.

The simplest, and by far most popular, model capturing the essential physics of these systems is the periodic Anderson model (PAM) [11]. It describes conduction electron in extended band states interacting with more localized orbitals from the inner $f$ or $d$ shells. The possibility of a "hybridization gap" arises naturally as a result of the mixing of localized and extended electron states. The effect of the strong Coulomb repulsion $U$ is primarily to renormalize the hybridization parameters of the easily solvable $U = 0$ Hamiltonian. As a consequence, the gap is rescaled to a low value, typically in the 10 meV range, of the order of the characteristic energy $E^*$ of the quasiparticle band. An alternative approach, valid in the limit of strongly localized $f$ states, is the so-called Kondo lattice (KL) model, which describes the system as an array of local moments interacting with conduction-electron spins via a Heisenberg-type exchange interaction. A "spin-liquid" phase is predicted to occur at half filling (i.e. one conduction electron per localized spin), when every localized spin forms a singlet with a conduction electron without having overlap with singlets at neighbouring sites. Again one obtains an insulating ground state which can serve as a paradigm of the strongly correlated semiconducting phases formed at low temperature in real Kondo insulators. Correlation effects are maximised at large $U$ (small $J$) in the PAM (KL) model, respectively.

As with their metallic counterparts, the properties of those strongly correlated insulators, often termed "Kondo insulators" or "heavy-fermion semiconductors", involve an interplay between charge and magnetic (spin + orbital) degrees of freedom. In the large-$U$ limit of the PAM, "spin-charge decoupling" may actually occur, leading to different functional dependences on the Coulomb repulsion $U$ for the charge-and spin gaps. It is therefore essential to obtain detailed experimental information both on the electronic structure, using transport, optical, or photoemission techniques, and on the magnetic response, using conventional magnetometry and inelastic neutron scattering.

$YbB_{12}$ is of particular interest for such a study because it crystallizes in a very simple structure (NaCl type, with Yb cations and $B_{12}$ dodecahedra forming two interpenetrating $fcc$ sublattices). Also, the fact that the Yb electronic configuration is very close to $4f^{13}$ [12], which corresponds to a single-hole occupancy of the $4f$ shell, is a significant simplification for comparison with theories. The opening of a gap upon cooling below $T \sim 60$ K is now well documented from numerous measurements of the resistivity [13,14], specific heat [15,16], optical conductivity [17], photoemission [18], electron tunneling [19], and bolometric response [20] in this material. However, values obtained from different techniques may differ significantly, sometimes by a factor of 2 or more. Electrical transport measurements [14] estimate a gap of about 12 meV whereas the "onset" in the optical conductivity $\sigma(\omega)$ is located around 25 meV, followed by a continuous increase up to 40 meV [17]. It has been argued in [17] that this difference can happen because optical measurements probe the *direct* band gap which, in the case of a hybridised band structure, may be much larger than the energy of indirect transitions seen by other techniques. Another source of inconsistency between different estimates could be the existence of additional structures within the gap, as suggested in [20] and [21].

The first neutron time-of-flight study of the dynamical magnetic response in YbB$_{12}$ was carried out by Bouvet and coworkers [22] back in the early 1990s. The results reported in [23] showed clear evidence for a spin-gap (defined as the energy threshold below which no magnetic signal can be detected) of about 10 meV in the low-temperature magnetic excitation spectra, with a rather complex peak structure just above the gap edge. In a subsequent study performed with improved experimental conditions [24], we could confirm the existence of the gap, and work out a more complete and accurate picture of the excitation spectrum and its temperature dependence. The key features are the complete lack of a quasielastic signal at $T =$ 10 K, implying that the ground state is a singlet (result confirmed subsequently by measurements with a higher resolution [12]), and the existence of three distinct magnetic excitations, two sharp peaks at 15 and 20 meV, and a much broader one at about 38 meV. Upon heating, a quasielastic component is seen to grow at low-energies whereas the inelastic peaks at 15 and 38 meV are strongly suppressed. At 100 K and up, the spectrum essentially consists of a broad ($\Gamma/2 \sim 15$ meV) quasielastic peak, characteristic of a single-ion spin-fluctuation regime, but still contains a residual peak at about 23 meV, which does not seem to decay appreciably at least up to 160 K. Recently, the first neutron scattering experiments on a single-crystal have been reported by Iga et al. [25]. The main result of that work is the observation of a dispersive branch of excitations between 15 and 20 meV, which could account for the lower two peaks in the powder spectra. A more systematic investigation of the $Q$ and $T$ dependences of the various excitations is still needed.

The existence of a gap at low temperature in the magnetic excitation spectra is clearly a key property, which seems to single out YbB$_{12}$, together with other Kondo insulators, among more conventional strongly correlated systems showing a quasielastic response [26, 27]. However, one should be careful in assuming a direct connection between this spin gap and the gap observed in other physical properties. We have already noted that even in the simplest models (PAM, KL) capable of describing this class of materials, spin and charge gaps may become decoupled in the strong correlation limit. Experimentally, the value obtained by neutrons in YbB$_{12}$ (10–12 meV) is close to those derived from the magnetic term in the heat capacity [15, 16] or from transport [13, 14] measurements, but much smaller than the optical conductivity threshold [17]. The origin of the near-gap excitations has given rise to a variety of theoretical speculations. In [23], the peak at 15 meV was ascribed to a *s*-wave magnetic exciton corresponding to the excitation of one electron from the Kondo-singlet ground state to a *local* bound state of *s* symmetry around the Yb site [28]. The energy of this exciton is expected to be slightly less than $k_B T_K$. Another excitation to a *delocalised* band state could account for the second peak at 20 meV (of the order of $k_B T_K$ or higher). A similar starting point was adopted by Liu [29] who treats the insulating ground state in $YbB_{12}$ as a lattice of spin singlets, expressed as a linear combination of $f^2$ and $f^1d^1$ states (standing for the $f^{14}$ and $f^{13}d^1$ configurations in the real system), in which the gap arises from local effects. In this model, the peak structure in the neutron spectra is ascribed to transitions into excited states which are represented phenomenologically by two sets of narrow conduction bands formed by *d-f* hybridization.

Riseborough [30], on the other hand, emphasizes the role of antiferromagnetic (AFM) correlations in the dynamical magnetic response. Starting from a slave-boson, mean-field treatment of the PAM, he introduces spin-dependent interactions between quasiparticles by calculating higher-order terms in a $1/N$ expansion. The important result is that, if magnetic correlations become strong enough, a soft branch of sharp, paramagnonlike excitations can form below the threshold of electron-hole excitations. It is argued in [30] that this mechanism could account for the low-energy structure in the $YbB_{12}$ spectra.

In this paper, we present a detailed study of how the magnetic response described above for pure $YbB_{12}$ is modified by the introduction of nonmagnetic sites on the rare-earth sublattice. Lutetium was chosen for this substitution because it is practically isovalent to Yb in this structure and produces only a marginal change in the lattice constant. Information on the alloying behaviour is important because, in the band picture, the formation of the gap requires the coherent hybridisation of *f* orbitals with conduction band states, and thus can be expected to be relatively sensitive to the disruption of the periodicity on the *f* sublattice. If, on the other hand, Kondo singlets are formed quasi independently at each Yb site, the effect of substituting Lu should be much less pronounced. We also show that the dilution of the magnetic sites produces significant changes in Yb-Yb intersite magnetic correlations which can be revealed, even for powder, by a careful analysis of the $Q$ dependence of the magnetic spectra. Finally, we present an estimate of the magnitude of the crystal-field (CEF) splitting, which is an important parameter for understanding the formation of the singlet ground state. As the complex magnetic response makes it impossible to get an answer directly from the neutron spectra of pure $YbB_{12}$, $Er^{3+}$ ions substituted for Yb are used as local probes to obtain an order of magnitude of the CEF effect.

## 2. Experiments

Five powder samples of $YbB_{12}$ ($m$ = 7.97 g), $Yb_{0.75}Lu_{0.25}B_{12}$ ($m$ = 12.83 g), $Yb_{0.1}Lu_{0.9}B_{12}$ ($m$ = 13.1 g), $Yb_{0.9}Er_{0.1}B_{12}$ ($m$ = 8.87 g), and $LuB_{12}$ ($m$ = 5.85 g) have been synthesized by borothermal reduction of ytterbium and lutetium oxides and their mixtures at 1700°C under vacuum at the NASU Institute for Problems of Material Science (Kiev). In order to reduce the neutron absorption cross-section, all compounds were prepared using isotopically enriched boron (99.5% $^{11}B$). Accordingly, transmissions were found to be about 70% for $E_i$ = 80 meV and 60% for $E_i$ = 40 meV. Contamination by foreign phases was estimated from x-ray analysis to be always less than 2%. The room temperature lattice constants of $LuB_{12}$ and the alloys are within 0.1 % of that for pure $YbB_{12}$, $a_0$ = 7.4684(2) Å.

Neutron scattering experiments were carried out on the time-of-flight spectrometer HET at ISIS/RAL (UK) with two different incident neutron energies $E_i$ = 80 meV (choppers "*B*" and "*S*") and 40 meV (chopper "*S*"). Combining those data sets, one can obtain information on both the high-energy (15 ≤ $E$ ≤ 65 meV) and low-energy ($E$ ≤ 15 meV) spectral ranges.

The detectors were grouped into seven banks with average scattering angles $\langle 2\theta \rangle$ equal to 5°, 11.5°, 16.5°, 21.5°, 26.5°, 115°, and 133°. The flight path to the detectors was 4 m at $\langle 2\theta \rangle$ = 5° and 115°, and 2.5 m at $\langle 2\theta \rangle$ = 11.5° to 26.5° and 133°. The resulting energy resolutions (full width at half maximum of the elastic peak in a vanadium spectrum) at the incident neutron energy $E_i$ = 80 meV using chopper "*B*" [chopper "*S*"] were 2.15 meV [3.2 meV] for $\langle 2\theta \rangle$ = 5°, and 3 meV [4.7 meV] for $\langle 2\theta \rangle$ = 11.5° to 26.5°. The resolutions at $E_i$ = 40 meV (chopper "*S*" only) were 1.7 meV for $\langle 2\theta \rangle$ = 5° and 2.9 meV for $\langle 2\theta \rangle$ = 11.5° to 26.5°. Spectra were recorded at a total proton current of typically 1000 $\mu$Ah.

Measurements were performed on $Yb_{1-x}Lu_xB_{12}$ and $Yb_{0.9}Er_{0.1}B_{12}$ at selected temperatures in the range 10–172 K. Reference spectra for $LuB_{12}$ were recorded at $T$ = 15 and 159 K ($E_i$ = 80 meV), and at $T$ = 10 and 153 K ($E_i$ = 40 meV). Existing data for $YbB_{12}$ have been supplemented by new measurements at $T$ = 186 K and 250 K ($E_i$ = 80 meV), and at $T$ = 6 K and 46 K ($E_i$ = 40 meV). The latter spectra will not be discussed in the following since the low-energy (quasielastic) magnetic response in $YbB_{12}$ has been the subject of a previous report [12]. Here those data were used only for comparison with $Yb_{1-x}Lu_xB_{12}$. Absolute calibration of the spectral data was achieved by normalization to a vanadium standard.

## 3. Results

### 3.1. Separation of the magnetic component

To determine the magnetic component from the measured spectra, one needs to subtract out a background contribution arising mainly from nuclear scattering by phonons within the sample, but also in some cases from parasitic scattering from the sample environment. Spectra for the nonmagnetic reference material $LuB_{12}$ cannot be directly used to estimate the nuclear background at a given angle, especially in the presence of multiple scattering. As shown by Murani [31], a more reliable procedure consists in deriving the nuclear contribution at low scattering angles from that measured at high scattering angles on the *same* sample by applying an energy-dependent scaling factor. This factor is obtained, for each scattering angle, from independent measurements on the nonmagnetic reference compound. The method was used successfully in our previous study of $YbB_{12}$ [24], and it remains applicable here for energy transfers above ≈ 10 meV. Basic conditions are that the sample and the reference should have similar scattering and absorption parameters (true of the present materials), and also *comparable masses*. Here we worked on a $LuB_{12}$ sample with a mass much smaller than that of all other samples. This difference turned out to hamper the subtraction of nonmagnetic scattering in the low-energy region ($E$ ≤ 10 meV) of the spectra measured with $E_i$ = 40 meV, where a peak is observed on the energy-loss side of the incoherent peak. Its origin is parasitic scattering from the Al walls of the cryogenerator thermal shield, which affects the low- and high-angle spectra in different ways. The main process for low-angle spectra corresponds to double coherent elastic back-scattering, first on the sample, then on the thermal shield, or vice-versa. In both cases a peak is predicted to occur at 3.2 meV (2 meV) for a flight path from sample to detector of 2.5 m (4 m) respectively, with an intensity proportional to the coherent scattering cross section of each compound, in agreement with the experimental spectra. At high angles, scattering from the Al walls contributes directly (no scattering on sample involved) in the form of a peak on each side of the elastic line. Based on these observations, we have estimated the nonmagnetic background at low energies ($E \leq 10$ meV) in the low-angle spectra by scaling the corresponding $LuB_{12}$ data according to the ratio of the coherent scattering cross sections and masses for each sample. For $T = 10$ and 153 K, the measured $LuB_{12}$ spectra could be used directly. For intermediate temperatures, the temperature dependent part of the nonmagnetic contribution was interpolated assuming simple Bose factor dependence. In the spectra measured at $E_i = 80$ meV, no significant parasitic signal was detected in the foot of the elastic peak, possibly because of a reduction in Debye-Waller factors, and also because the resolution is poorer; the standard phonon correction procedure was therefore applied as described above.

*3.2. Magnetic response*

Magnetic spectra for different compositions, temperatures, and scattering angles have been obtained from the experimental spectra (corrected for self-screening, and normalized to a vanadium standard) by subtracting out the calculated nuclear component as explained in the preceding section. All results[5] have been scaled to $Q = 0$ according to the calculated $Yb^{3+}$ single-ion magnetic form factor [32]. In figures 1 and 2 the magnetic spectral response of $Yb_{0.75}Lu_{0.25}B_{12}$ measured at incident energies $E_i = 80$ and 40 meV respectively is compared with that of pure $YbB_{12}$ (data for $E_i = 80$ meV are from Ref. [24]). The results indicate that the two sets of data remain very similar[6]. At the lowest temperature ($T = 10$ K), it is particularly remarkable that the gap found in $YbB_{12}$ is *not* suppressed by the substitution of 25 percent of Yb by Lu (figures 1 and 2). The measurements with the higher resolution ($E_i = 40$ meV) also show that, as reported previously for $YbB_{12}$ [12], no significant quasielastic magnetic scattering exists. On the other hand, the low-energy peak structure just above the gap is modified. In figures 1 and 2, the $Yb_{0.75}Lu_{0.25}B_{12}$ spectra have been fitted using the same Lorentzian components as for $YbB_{12}$ in Ref. [24], namely three inelastic peaks at 15, 20, and 38 meV, plus an extra quasielastic line appearing in the spectra only at temperatures higher than 40 K. The parameters from the fits are displayed in figures 3 (peak energies), 4 (line widths), and 5 (intensities), together with the previous data for $YbB_{12}$. Note that, in the latter plot, the "intensity" shown for each spectral component $i$ is *not* the spectral weight in the usual sense but the value of the Kramers-Kronig integral

$$\int_{-\infty}^{+\infty} S_i(Q=0, E, T) \left[1 - \exp(-E/k_B T)\right] (1/E) \, dE, \tag{1}$$

---

[5] The same treatment has been applied also to the $Yb_{0.9}Er_{0.1}B_{12}$ spectra since the $Q$-dependences of the magnetic form factors for $Yb^{3+}$ and $Er^{3+}$ do not differ significantly)

[6] For $E_i = 80$ meV (figure 1), one notes a higher intensity in the energy gain region of the low-temperature spectra. This extra scattering, which cannot correspond to any physical process within the sample, may be an artefact of the nonmagnetic background correction in this energy region where both phonon and magnetic contributions become exceedingly small. Accordingly, the effect is largest in the most diluted sample (see figure 6 below), whereas it appears to be reduced in the spectra measured at lower incident energy (figure 2) for which a more reliable procedure has been applied (section 3.1).

which expresses the partial contribution $\chi_i$ to the static susceptibility in terms of the measured scattering function $S_i$ (see Section 4 below). In this way the effect of the detailed-balance factor is removed and the $T$ dependence of the underlying physical mechanism is more directly apparent.

Rather unexpectedly, the modification in the lower two peaks seen on the spectra appears to reflect an increase in their line widths, rather than a reduction of their spectral weight, which remains roughly unchanged. The peak at 40 meV does not appear to be observably affected by Lu substitution.

As a whole, the temperature dependence of the magnetic spectra for $Yb_{0.75}Lu_{0.25}B_{12}$ is very similar to that reported previously for $YbB_{12}$. The main evolution is the gradual replacement of the low-temperature three-peak structure by a broad quasielastic response (width $\Gamma/2 \approx 15$ meV), with a residual inelastic signal centred at about 23 meV. We have checked the magnetic response in pure $YbB_{12}$ up to 250 K and found no significant change with respect to the spectrum at 159 K reported previously, apart from a small difference in the neutron energy gain region, which simply reflects the increase in the temperature factor. This shows that the magnetic spectral response of $YbB_{12}$ becomes temperature independent in this range and that the inelastic peak at 20 meV is not suppressed at least up to room temperature.

In order to study the effects of high dilutions, approaching the single magnetic impurity limit, we have measured the magnetic response in a sample of $Yb_{0.1}Lu_{0.9}B_{12}$ (incident energy $E_i = 80$ meV). The spectra for $T = 80$ and 160 K do not differ significantly. They were therefore added up so as to improve statistics (the resulting spectra will be denoted by their average temperature of 120 K). For the same purpose all detectors covering scattering angles from 11.5° to 26.5° have been grouped together. The result is displayed in the lower frame of figure 6, with the spectrum of $YbB_{12}$ at $T = 105$ K also shown for reference. One sees that the $Yb_{0.1}Lu_{0.9}B_{12}$ exhibits the same featureless quasielastic profile found in pure $YbB_{12}$. This is consistent with the view that the properties above 100 K are dominated by single-ions effects. The important point is that, even in the dilute limit, one does not recover the usual narrow-peak structure expected from well-resolved crystal-field excitations in normal rare earths.

The spectrum for $T = 20$ K (upper frame in figure 6) consists of a broad signal above 30 meV, whose spectral shape cannot be precisely defined within the accuracy of the data. No evidence remains for the existence of low-energy peaks but, most surprisingly, no quasielastic scattering could be detected at low energies. The persistence of a "spin-gap response" in a concentration range where clear metallic behaviour has been reported for a number of other physical properties is puzzling. It might indicate a decoupling between charge and spin degrees of freedom as reported previously for the mixed-valence compound CeNi [33]. One should keep in mind, however, that Yb scattering is quite weak in the present alloy, and the subtraction of the phonon background thus entails significant errors (as high as 50%) in the resulting magnetic signal. In particular, it is not possible to decide whether the apparent loss of overall intensity in comparison with pure $YbB_{12}$ results from a transfer of spectral weight to a different energy window, or simply from systematic errors in the background correction. Attempts to analyse the low-energy region with better resolution using incident neutron energy $E_i = 40$ meV were hindered by spurious contaminations of the type described in section 3.1, which become intractable when magnetic scattering is low. It would be important in the future to confirm the present results on a less diluted compound.

At this point, we have shown that the substitution of Yb by Lu results in the broadening, then in the gradual suppression, of the two near-gap peaks in the magnetic spectral response of $Yb_{1-x}Lu_xB_{12}$ at low temperature. On the other hand, the "spin gap" is not suppressed by the dilution process. The peak at $E \approx 40$ meV also has dominant single-ion character. Above $T \sim 100$ K all compounds exhibit a broad quasielastic response, with an extra component around 20 meV in the Yb-rich limit. The role played by crystal-field (CF) interactions remains unclear for lack of any detectable transitions in the measured spectra. In the next section, we present an attempt to shed light on this question by using dilute Er ions as sensors of the CF potential in $YbB_{12}$.

*3.3. Crystal-field excitations in $Yb_{0.9}Er_{0.1}B_{12}$*

Measurements performed on a $Yb_{0.9}Er_{0.1}B_{12}$ sample are shown in figure 7. To extract the magnetic signal associated with $Er^{3+}$ ions, $Yb_{0.75}Lu_{0.25}B_{12}$ magnetic spectra, scaled to the actual Yb content, were used as a reference. For $T = 24$ K (upper frame), the spectrum consists of three distinct peaks at energies of approximately 12, 17, and 24 meV. These energies do not change with increasing temperature (figure 7). At $T = 169$ K the amplitude of the peak at 17 meV decreases by a factor of about two without significant broadening, whereas the intensity of the peak at 12 meV remains nearly unchanged. The temperature evolution of the third peak at 24 meV is more uncertain because of low statistics and, in addition, double scattering from the 12-meV excitation may be significant. From the positions of the peaks and their spectral weight at $T = 24$ K, a unique set of cubic CEF parameters can be derived for $Er^{3+}$: $B_4^0 = (-46 \pm 32) \times 10^{-6}$ meV, $B_6^0 = (-4.0 \pm 0.2) \times 10^{-6}$ meV. The high-temperature spectrum was then used for a check of consistency. Calculated spectra are shown together with the experimental data points in both frames of figure 7, and the corresponding level scheme is sketched in figure 8.

One sees that a single set of CEF parameters describes the experimental spectra quite well both at low and high temperature. The obtained level scheme accounts for the difference in the temperature evolution of the 12- and 17-meV peaks. The decrease in the intensities of the $\Gamma_8^{(3)} \to \Gamma_8^{(2)}$ and $\Gamma_8^{(3)} \to \Gamma_6$ transitions reflects the thermal depopulation of the ground state. However, the intensity of the experimental peak at 12 meV does not decrease substantially because temperature increases the occupation of the $\Gamma_8^{(2)}$ excited state, causing the appearance of the transition between $\Gamma_8^{(2)}$ and $\Gamma_8^{(1)}$ with a nearly equal energy of 11.8 meV (longer dashed arrow in figure 8). It can also be noted that, according to the relative values of the magnetic dipole transition matrix elements, most of the magnetic cross section of the $Er^{3+}$ ions goes to quasielastic scattering (see dashed curve in figure 7).

The next step is to estimate the CEF splitting of $Yb^{3+}$ in $YbB_{12}$ on the basis of the $Er^{3+}$ results. The calculation relies on the assumption that effect of the CEF potential can be scaled from one rare-earth to another in the same local environment by factorizing the parameters $B_4^0$ and $B_6^0$ in the form $B_n^0 = \theta_n \langle r_{4f}^n \rangle A_n^0$ [34], where the tabulated constants $\theta_n$, and the average $\langle r_{4f}^n \rangle$ of the $n^{th}$ power of the 4*f* shell radius both characterize the *element* under consideration, whereas $A_4^0$ and $A_6^0$ parameterize the contribution of the *local charge distribution* to the CEF potential. This is clearly an approximation, which has been found to hold rather well in the $R$Al$_3$, $R$Pd$_3$, and (in so far as data exist) the $R$B$_6$ series, but failed to account even for the sign of the $B_n^m$ parameters in the $R$Ni$_5$ series, possibly because of larger $d$-electron contributions. In dodecaborides, there is not enough experimental information available to be sure that the approximation is valid, but existing data for heavy rare earths [35] suggest that it could be the case.

Using the values of $B_4^0$ and $B_6^0$ determined above for Er$^{3+}$ in Yb$_{0.9}$Er$_{0.1}$B$_{12}$ (or, equivalently, $A_4^0 = -12 \pm 8$ meV and $A_6^0 = -22.1 \pm 1.1$ meV), the corresponding crystal field parameters for Yb$^{3+}$ in YbB$_{12}$ can be readily calculated by substitution of the appropriate values of $\theta_n \langle r_{4f}^n \rangle$ for the lower Yb$^{3+}$ spin-orbit multiplet $J = 7/2$. One obtains $B_4^0 = 0.09 \pm 0.06$ meV and $B_6^0 = -0.28 \pm 0.02$ meV, which leads to the CEF level scheme depicted in figure 8. The ground state is a $\Gamma_8$ quartet and the two excited doublets $\Gamma_7$ and $\Gamma_6$ lie at energies of $6.0 \pm 1.5$ and $11.2 \pm 0.6$ meV, respectively. Keeping in mind the above-mentioned caveats, these values should be taken only as order of magnitudes of the CEF effect in YbB$_{12}$ since hybridisation effects can change the total CEF splitting by as much as a factor of two, as observed experimentally in the Ce(La,Nd)Ni series of compounds [36]. However, they lend support to the idea that the CEF is a relevant energy scale in the low-temperature properties of this compound.

## 4. Discussion

The data presented above reveal that the main effect of substituting Yb by Lu in Yb$_{1-x}$Lu$_x$B$_{12}$ is a broadening, followed by the disappearance, of the two near-gap peaks in the magnetic spectral response at low temperature. The "spin gap", on the other hand, is not suppressed by the dilution process. Above $T \sim 100$ K a broad quasielastic response is recovered for all concentrations, with an extra component around 20 meV in the Yb rich limit. These results have important implications for understanding the ground state and low-lying excitations in YbB$_{12}$. In particular, it can be inferred that *i)* coherence on the Yb sublattice and/or magnetic ("intersite") correlations strongly affect the low-energy magnetic excitations, and *ii)* the gap in the magnetic excitation spectra arises, at least partly, from single-site (local) processes.

Concerning the first point, further insight can be gained from a comparison of the dynamical susceptibilities in pure and Lu substituted YbB$_{12}$ using Kramers-Kronig analysis. In our previous paper [24] we had reported the existence of a considerable discrepancy below $T \sim 50$ K (see figure 9) between the bulk magnetic susceptibility $\chi_{\text{bulk}}$ as measured in a magnetometer [16] and the local static susceptibility $\chi_{\text{loc}}$ derived from the neutron scattering spectra using the standard Kramers-Kronig relation

$$\chi_{\text{loc}}(T) = K \int_{-\infty}^{+\infty} S(Q=0,E,T)\left[1 - \exp(-E/k_B T)\right] (1/E) \, dE, \qquad (2)$$

where $K$ is a prefactor equal to 0.22 if $S$ is expressed in units of barn meV$^{-1}$ sr$^{-1}$ / f.u., and $\chi_{\text{loc}}$ in emu/mole. The discrepancy was attributed to the fact that the $Q$ dependence of the scattering function used for extrapolating the measured data to $Q = 0$ was taken to follow the single-ion Yb$^{3+}$ form factor. This assumption obviously ignores the extra $Q$ dependence arising from intersite correlations, which seems to play an important role in pure YbB$_{12}$. On the other hand, the temperature dependences of $\chi_{\text{bulk}}$ and $\chi_{\text{loc}}$ in Yb$_{0.75}$Lu$_{0.25}$B$_{12}$ turn out to coincide within error bars, with no appreciable discrepancy even below 50 K. This suggests that the reason why the local susceptibility calculated from the neutron spectra fails to reproduce the low-temperature drop of $\chi_{bulk}(T)$ in YbB$_{12}$ is that the scattering function measured on powder at finite scattering angles is enhanced by antiferromagnetic correlations, and that its reduction for $Q \rightarrow 0$ is not properly represented by using the single-ion form factor. The absence of such pronounced discrepancy in the alloy system probably traduces the fact that the magnetic intensity is much more distributed in reciprocal space. It is quite remarkable that, if one performs the Kramers-Kronig calculation on a partial YbB$_{12}$ spectrum ($T$ = 10 K) in which the low-energy peaks have been subtracted out, the reduction in the resulting value for $\chi_{loc}$ (diamond marker in figure 9) brings it very close to the experimental $\chi_{bulk}$. In other words, the spectrum with the near-gap peaks removed probably yields a rough approximation to the magnetic response near $Q = 0$. This observation lends additional support to the idea that the low-energy region is dominated by magnetic correlations. Triple-axis experiments on single-crystals are now in progress to confirm this interpretation.

AFM correlations have long been known to play a prominent role in metallic heavy fermion systems [37, 38], especially in the vicinity of quantum critical points. In Kondo insulators it has been argued, on the basis of a Kondo lattice model, that they can strongly renormalize the magnetic and transport gaps [39]. However, clear evidence for a $Q$ dependence in the magnetic response measured by inelastic neutron scattering has been reported so far only for SmB$_6$ [40] and for CeNiSn [41], a material with pronounced Fermi surface anisotropy in which the gap is thought to exist only on a fraction of the Fermi surface. In contrast, YbB$_{12}$ seems to provide a unique example of a Kondo insulator in which a well-identified spectral component can be associated with intersite correlations.

In this respect, our results are in line with expectations from Riseborough's model. The "spin exciton" obtained in Ref. [30] should primarily be observable at a wave vector $\boldsymbol{Q_0}$ close to the Brillouin zone corner where both the exchange interaction $J(\boldsymbol{Q_0})$ and the real part of the irreducible susceptibility $\chi_0(q,\omega)$ are large. The branch is expected to show dispersion and to merge into the continuum as $q$ deviates from $\boldsymbol{Q_0}$. The absence of lifetime broadening in the low-energy peaks, in comparison with the 38-meV peak, is explained by the fact that the corresponding excitations fall in the gap below the Stoner continuum. One can also understand why the energy of the lower peaks, associated with a wave vector spanning the Brillouin zone and thus connected with the value of an *indirect* gap, has to be smaller that the direct gap probed by the optical measurements. Recently, the model has been extended to treat the effect of Lu substitution [42], and it was found that the low-energy peaks are strongly suppressed, again in qualitative agreement with our data.

On the other hand, the remarkable robustness of the spin gap and of the 38-meV peak against Lu substitution is difficult to reconcile with the above picture, and rather seems to favour a single-site mechanism of the type assumed in [28] or [29]. Unfortunately, there is at present no realistic calculation based on these models which could allow a quantitative comparison to be made with experimental data. Even qualitatively, it seems unlikely, based on our estimate of the CEF splitting, that the fine structures predicted in [29] may be resolved, if they exist, within the broadening of 38-meV peak.

From an experimental viewpoint, it can be noted that the position of the higher excitation near 38 meV, which is little affected by Lu substitution but strongly suppressed by increasing temperature, roughly corresponds to the shoulder in the optical conductivity, which Okamura et al. [43] have found to exist, with no significant shift in its position, at concentrations up to $x_{Lu}$ = 0.5. In the optical data for YbB$_{12}$, the gradual disappearance of the gap at increasing temperature took place in the form of a gradual recovery of the conductivity at lower energies, tentatively ascribed to the appearance of mid-gap states. In our neutron data for $T \approx$ 10–20 K, the 38-meV peak has a full width at half maximum of about 20 meV, which means that its tail extends over the entire region where Okamura et al. [17] observe a steady increase in $\sigma(\omega)$. There is at present some controversy as to whether transitions across the *indirect* gap in the PAM might be observable in the optical conductivity [44]. If it is the case, the corresponding gap energy in YbB$_{12}$ might lie in the 40 meV region, rather than near 10 meV as assumed previously, reinforcing the view that the lower peaks in the neutron spectra may correspond to in-gap (excitonlike) modes. It should be mentioned in this connection that, both in neutron [40, 24, 33, 45, 46] and optical conductivity [47, 17, 48] measurements, striking similarities have been found between the energy response in insulating (SmB$_6$, YbB$_{12}$) or metallic (CeNi, YbAl$_3$,) correlated electron materials. While no satisfactory explanation of this experimental fact has been given so far, it should obviously be kept in mind when discussing the origin of the spin-gap and low-energy magnetic response of Kondo insulators.

For temperatures higher than ~100 K, the magnetic excitation spectra for pure YbB$_{12}$ and for the Yb$_{1-x}$Lu$_x$B$_{12}$ solid solutions coincide. This indicates that the high temperature state in YbB$_{12}$ is dominated by single-ion effects. However, the spectral shape still does not correspond to a simple spin-fluctuation regime as observed in many mixed-valence systems. In particular, we have seen that in YbB$_{12}$, the inelastic peak around 20 meV still exists at temperatures as high as 250 K. In view of the total splitting estimated in the previous section, this peak could be due to a transition between CEF levels. A difference between our estimate of the splitting and the actual peak position is not unexpected considering the approximations discussed above, and also because hybridisation effects are completely neglected in our simple calculation. But in any case the magnetic spectral response should contain, besides the quasielastic peak, *two* inelastic components of comparable intensities, corresponding to transitions between the levels of the split spin-orbit Yb$^{3+}$ ground state multiplet $J$ = 7/2 (figure 8). Only under very special conditions of parameters could those transitions possibly account for the single peak at 23 meV, and even less so if one considers that its width is only half of that of the quasielastic line.

Therefore, the CEF interaction alone cannot account for the inelastic peak in the high-temperature magnetic spectral response of YbB$_{12}$. Let us recall that even in the case of the most diluted compound Yb$_{0.1}$Lu$_{0.9}$B$_{12}$, the magnetic spectra could not be analysed as a set of CEF transitions. However, it may still be significant that the energy of the only remaining inelastic peak in the YbB$_{12}$ spectrum at $T > T_K$ and the Kondo temperature $T_K$ ~100 K derived from the width of the quasielastic peak [24], fall in the same energy range as the estimated value of total CEF splitting. It has been argued [49] in the case of another Kondo-lattice material, CeNiSn, that proper consideration of the interplay between Kondo effect and CEF interactions, with comparable energy scales, is necessary to explain the unconventional magnetic spectral response.

In YbB$_{12}$, however, one has to cope with the extra complication raised by the existence of a strong phonon peak at an energy of about 15 meV [24]. Whereas we are quite confident that the subtraction procedure applied above correctly separates the magnetic signal, the existence a weak residual nuclear signal cannot be completely ruled out without further polarized neutron experiments. This will be the subject of a forthcoming study.

## 5. Conclusion

In this neutron scattering study of $YbB_{12}$ solid solutions, we have shown that the disruption of periodicity on the Yb sublattice achieved by substituting nonmagnetic trivalent Lu specifically affects the low-energy peak structure near the gap edge. From a Kramers-Kronig analysis of the spectra, it can be inferred that at least the 15 meV peak, which is also rapidly suppressed by heating to ~ 100 K, results from Yb-Yb AFM correlations in the paramagnetic state. This makes $YbB_{12}$ the first likely example of a Kondo insulator in which a well-defined spectral component can be traced back to the effect of intersite couplings.

On the other hand, the formation of the spin gap in the low-temperature spectra was found to be rather insensitive to Lu substitution, suggesting that it originates from a predominantly local (incoherent) mechanism. The same holds for the broad peak around 40 meV, which is not too surprising since, as we have seen above, the temperature dependence of this peak closely follows that of the spin gap.

The data for the most diluted composition $x_{Lu} = 0.9$ indicate that a significant suppression of the magnetic signal at low energies (spin gap or pseudo-gap) still occurs deep inside the metallic regime. This result may seem puzzling at first sight, but it can be related to the previous observation of a broad *inelastic* response in several intermetallic mixed-valence compounds such as CeNi [33] or $YbAl_3$ [45] which, for the latter system, was proved to exist up to high levels of Lu substitution [46]. The possibility that the spin-gaplike spectral response in Kondo insulators and the inelastic response in mixed-valence intermetallics might have a common physical origin is an open and most interesting question.

At higher temperatures, the spin gap is gradually suppressed and the spectral response, dominated by incoherent, single-ion processes, becomes independent of the Lu concentration. However, even in diluted samples, one never recovers an ioniclike spectrum with well-defined CEF transitions. If such transition existed, our estimate of the CEF splitting from Er substitution experiments indicates that they should have been observed in our energy window. Instead of that, the magnetic response remains dominated by spin fluctuation processes, with a residual peak at about 20 meV. The latter feature is not yet fully understood, but it may give us a clue that some peculiarities of the magnetic response found experimentally may result from an interplay of distinct physical mechanisms (Kondo fluctuations, CEF, lattice vibrations) having comparable characteristic energies in the present material.

**Acknowledgements**

P A A, K S N and V N L are grateful to ISIS facilities for financial support and hospitality during they stay in RAL for measurements. We thank Y. Sidis for useful comments and suggestions. The work was supported by Grants No. 02-02-16521 (RFBR), 03-51-3036 (INTAS), HIII-2037-2003-2, and by the Russian State Program "Neutron Study of Condensed Matter"

**Figure captions**

Figure 1. Comparison of magnetic excitation spectra for $Yb_{0.75}Lu_{0.25}B_{12}$ and $YbB_{12}$ [24] at different temperatures measured with neutrons of incident energy $E_i$ = 80 meV; the average scattering angle of the detector banks included in the summation is $\langle 2\theta \rangle = 19$ degrees.

Figure 2. Comparison of magnetic excitation spectra for $Yb_{0.75}Lu_{0.25}B_{12}$ and $YbB_{12}$ [24] at different temperatures measured with neutrons of incident energy $E_i$ = 40 meV; the average scattering angle of the detector banks included in the summation is $\langle 2\theta \rangle = 19$ degrees.

Figure 3. Temperature dependence of the peak energies obtained from the fits of the $Yb_{0.75}Lu_{0.25}B_{12}$ and $YbB_{12}$ [24] spectra. Squares: $E_i = 80$ meV; circles: $E_i = 40$ meV.

Figure 4. Temperature dependence of the line widths (full width at half maximum) obtained from the fits of the $Yb_{0.75}Lu_{0.25}B_{12}$ and $YbB_{12}$ [24] spectra. Squares: $E_i = 80$ meV; circles: $E_i = 40$ meV.

Figure 5. Temperature dependence of the intensities (partial contributions $\chi_i$ to the static susceptibility derived from Kramers-Kronig integrals – see section 3.2) obtained from the fits of the $Yb_{0.75}Lu_{0.25}B_{12}$ and $YbB_{12}$ [24] spectra. Squares: $E_i = 80$ meV; circles: $E_i = 40$ meV.

Figure 6. Comparison of magnetic excitation spectra for $Yb_{0.1}Lu_{0.9}B_{12}$ and $YbB_{12}$ [24] at $T = 20$ and 120 K measured with neutrons of incident energy $E_i = 80$ meV; the average scattering angle of the detector banks included in the summation is $\langle 2\theta \rangle = 19$ degrees.

Figure 7. Magnetic excitation spectra of $Yb_{0.9}Er_{0.1}B_{12}$ at $T = 24$ and 169 K. Circles: experiment; solid lines: calculation using the set of CEF parameter ($B_4^0 = -46\times10^{-6}$ meV, $B_6^0 = -4.0\times10^{-6}$ meV) obtained from the fit; dashed line: same as previous, scaled by a factor of 10.

Figure 8. CEF level scheme for $Er^{3+}$ (a) and $Yb^{3+}$ (b) in $YbB_{12}$ using the parameters derived in the text; arrows in frame (a) correspond to the transitions observed in the experimental spectra.

Figure 9. Static magnetic susceptibility of Yb in pure and Lu-substituted $YbB_{12}$ obtained by bulk magnetization measurements [16] (solid and dashed lines) or calculated from the neutron spectra (circles and squares — see text). The diamond marker indicates the reduction (arrow) in the value obtained from the $YbB_{12}$ spectrum at $T = 15$ K if the spectral weight associated with the low-energy peaks is discarded. The dash-dotted line represents the $Yb^{3+}$ Curie law.

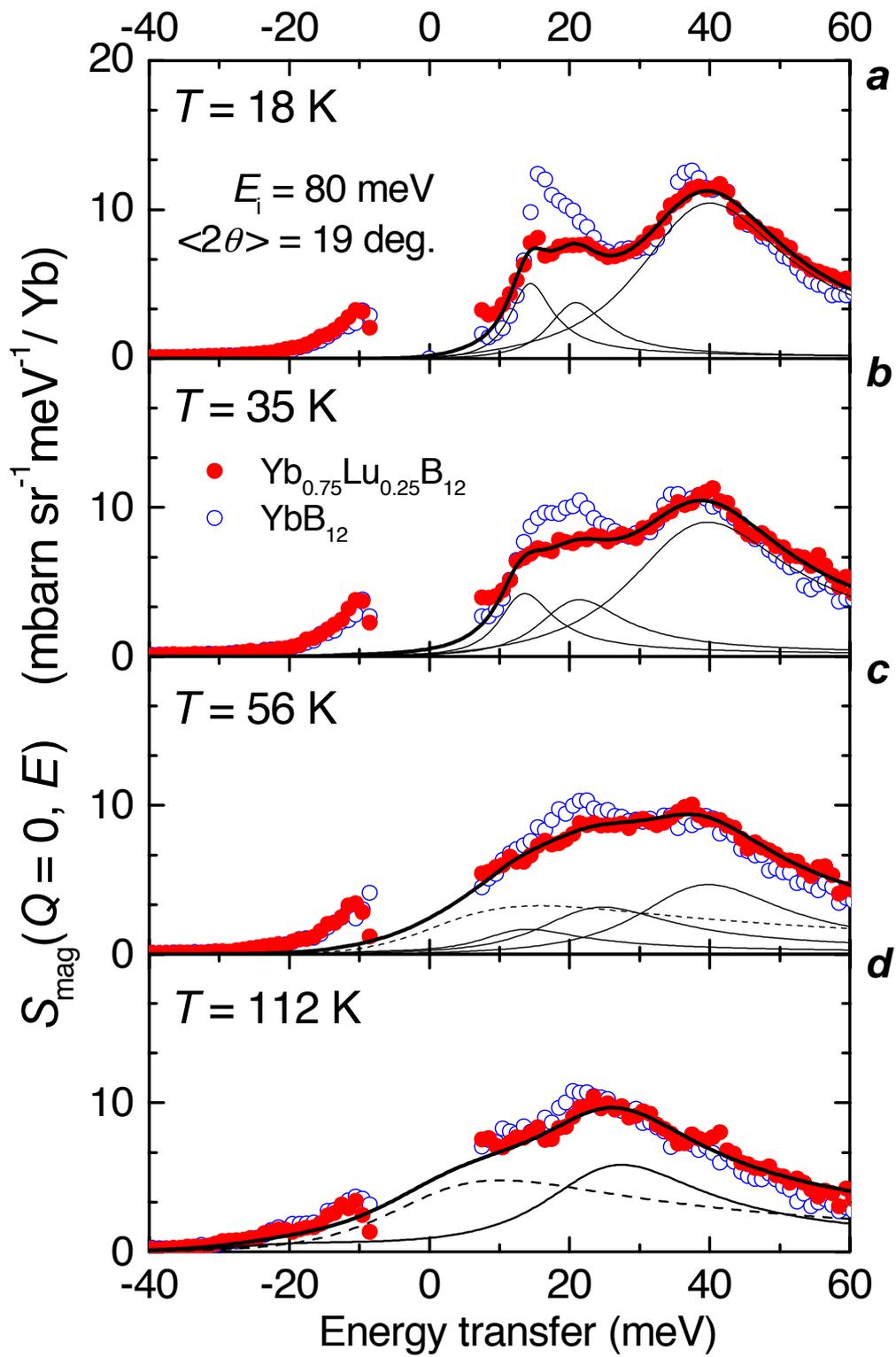

**Figure 1**

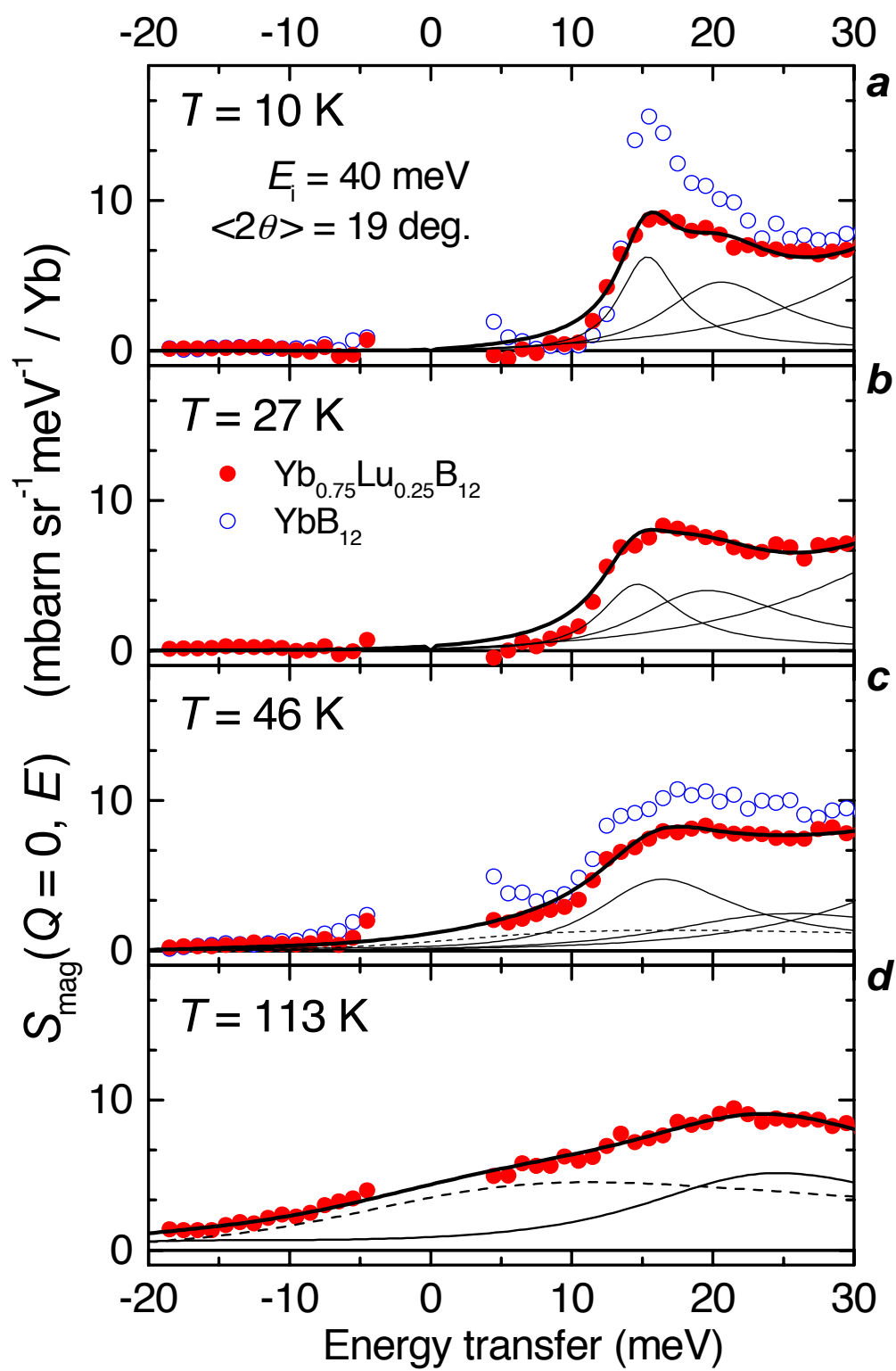

Figure 2

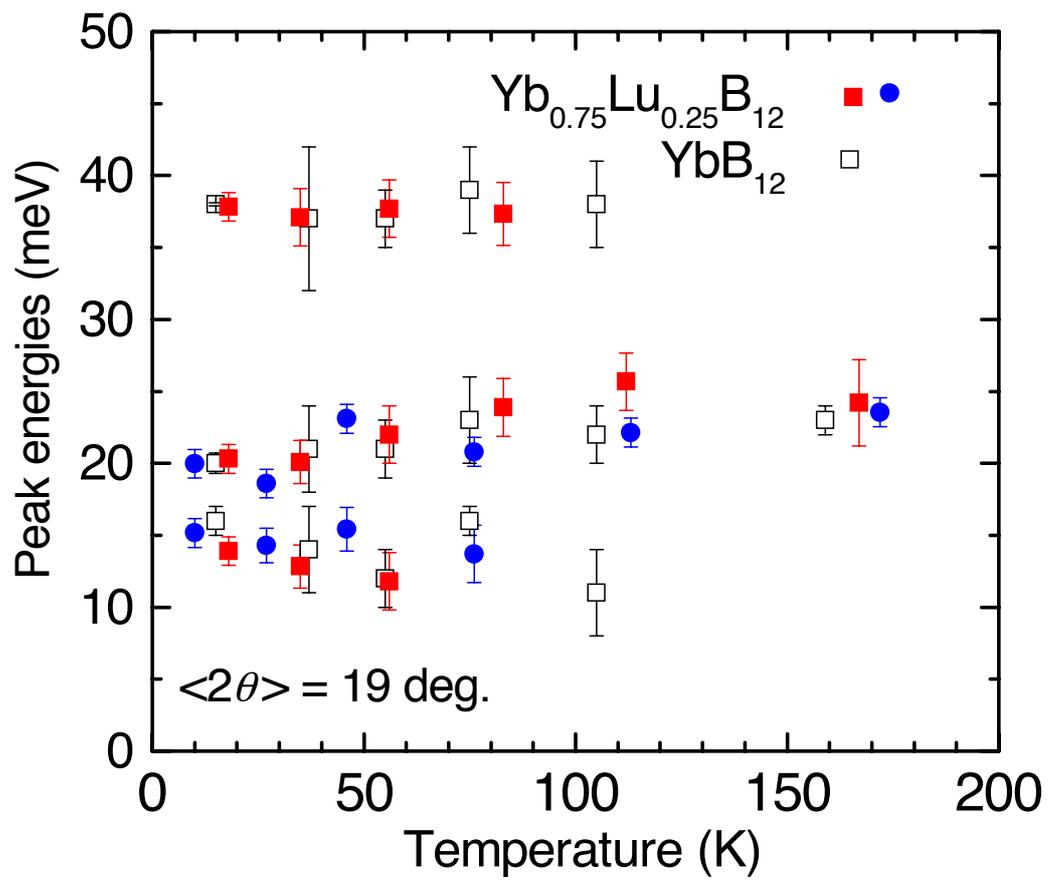

**Figure 3**

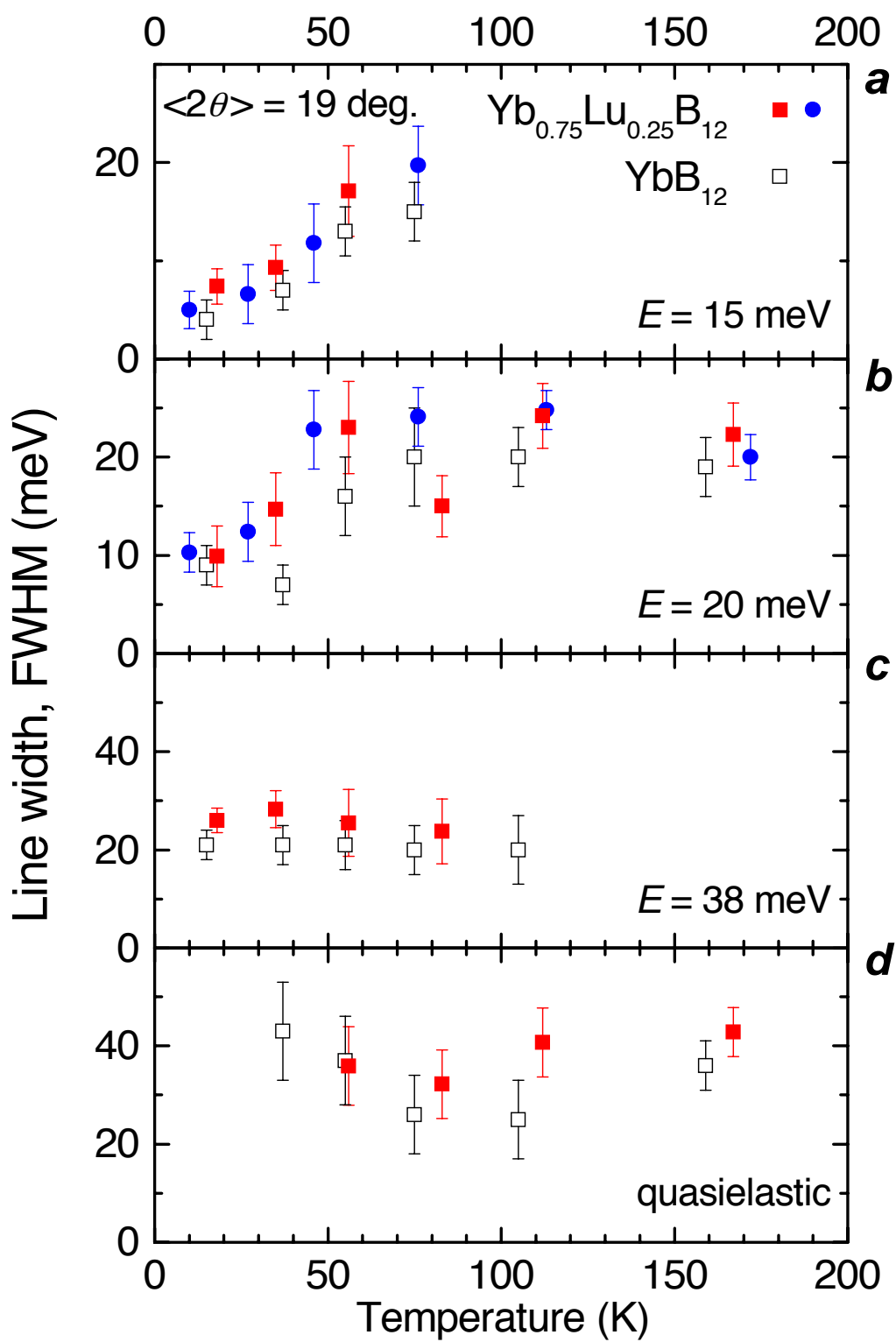

Figure 4

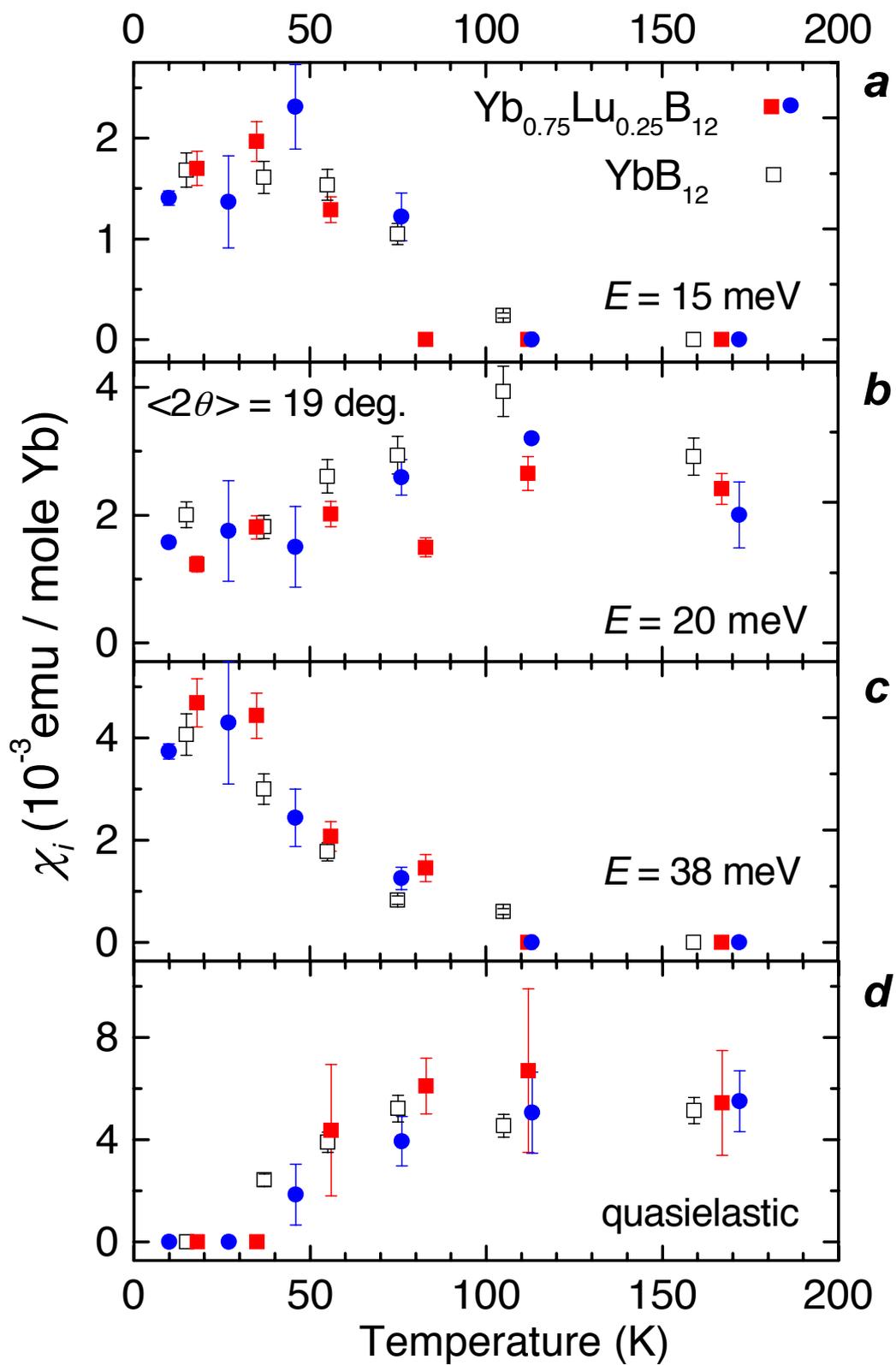

Figure 5

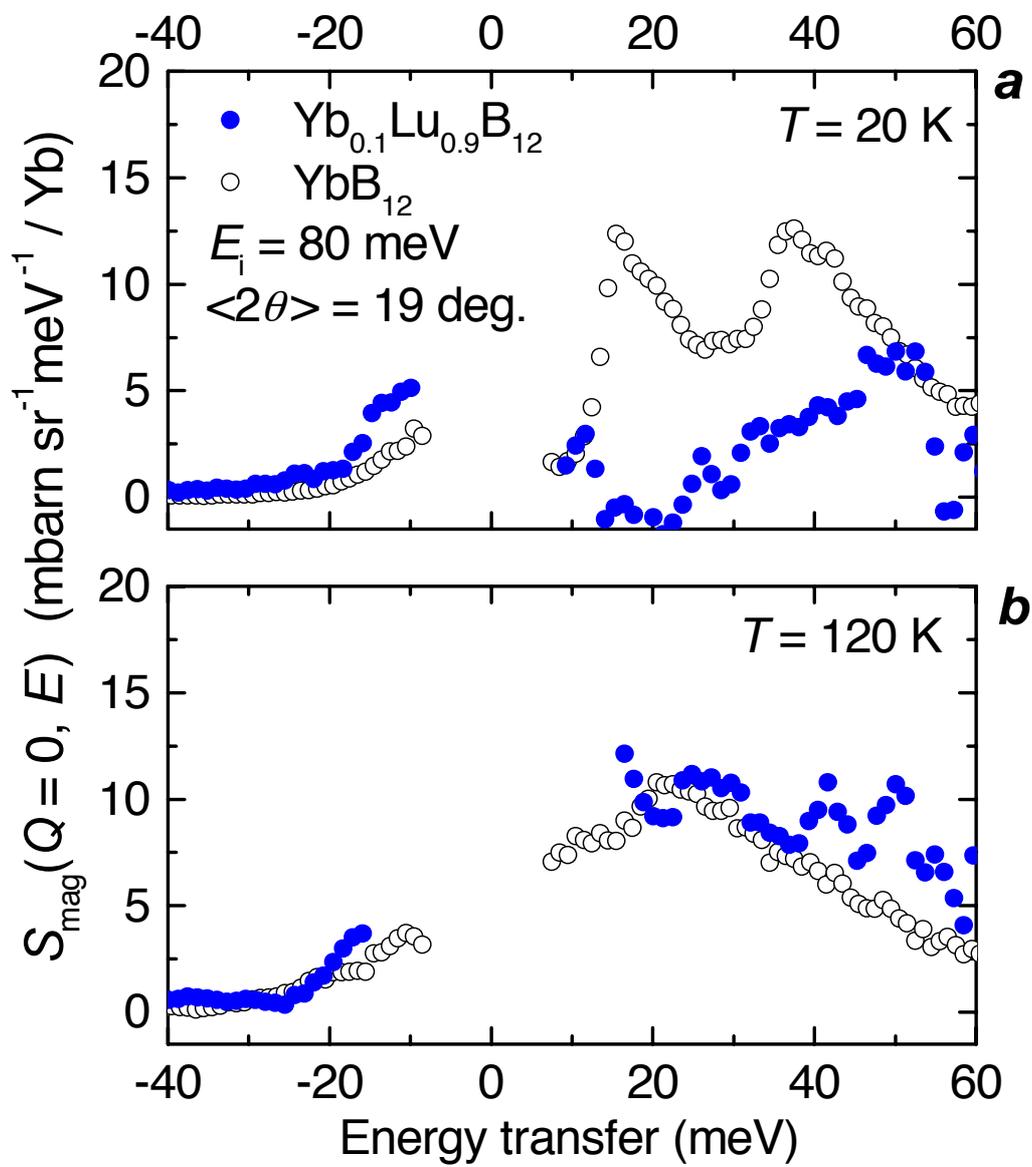

Figure 6

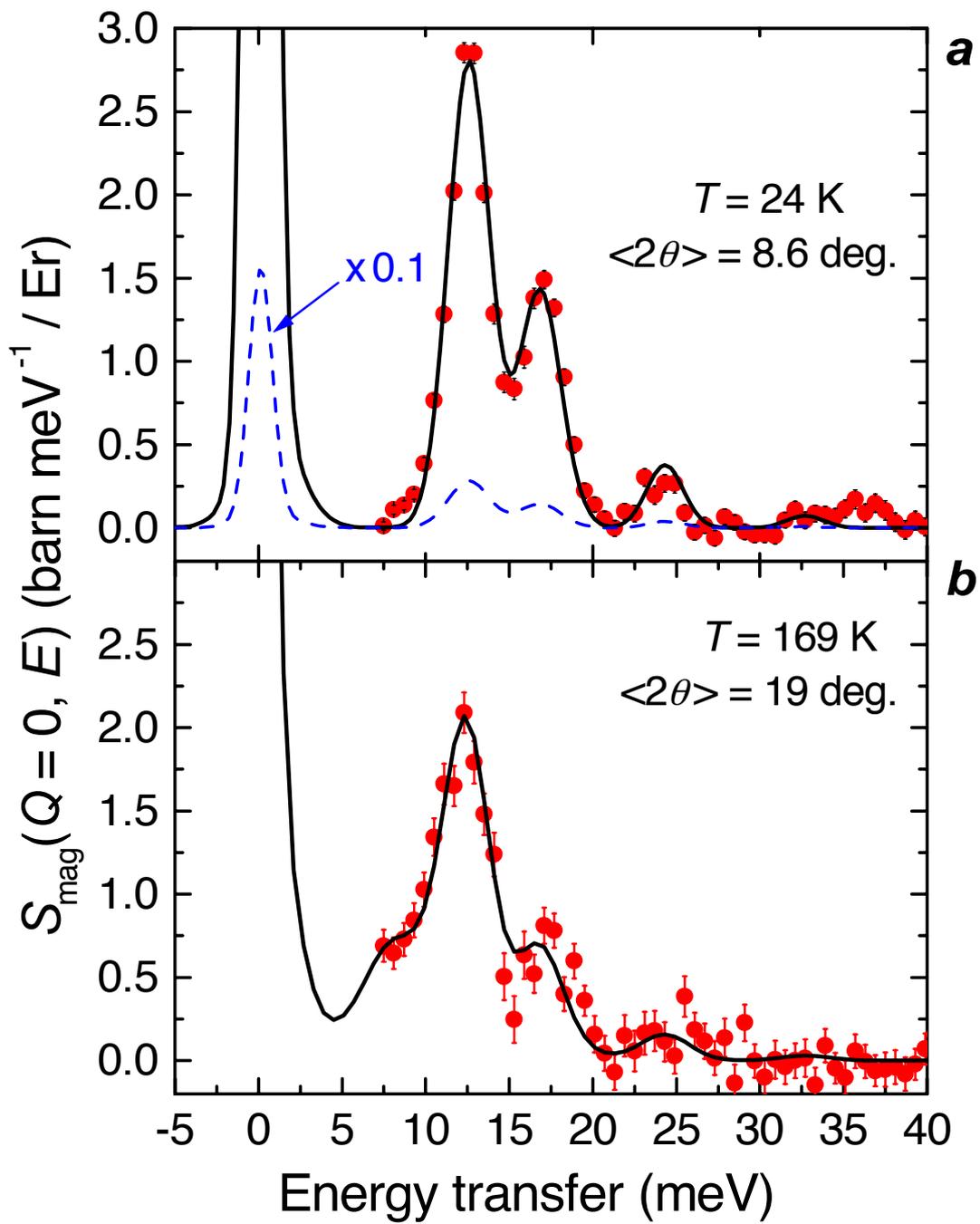

**Figure 7**

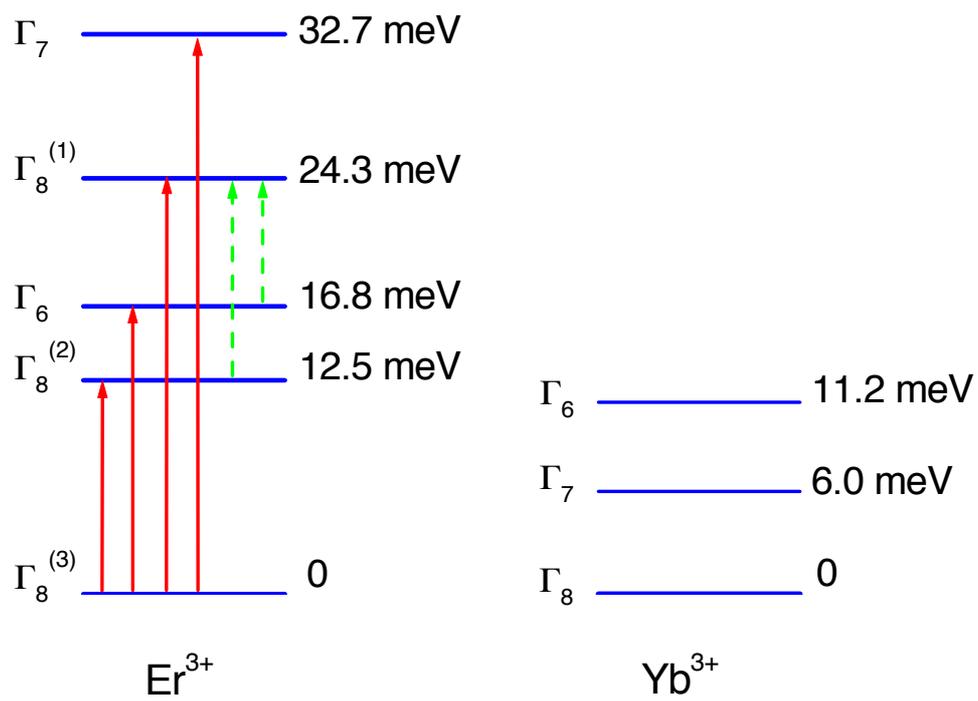

**Figure 8**

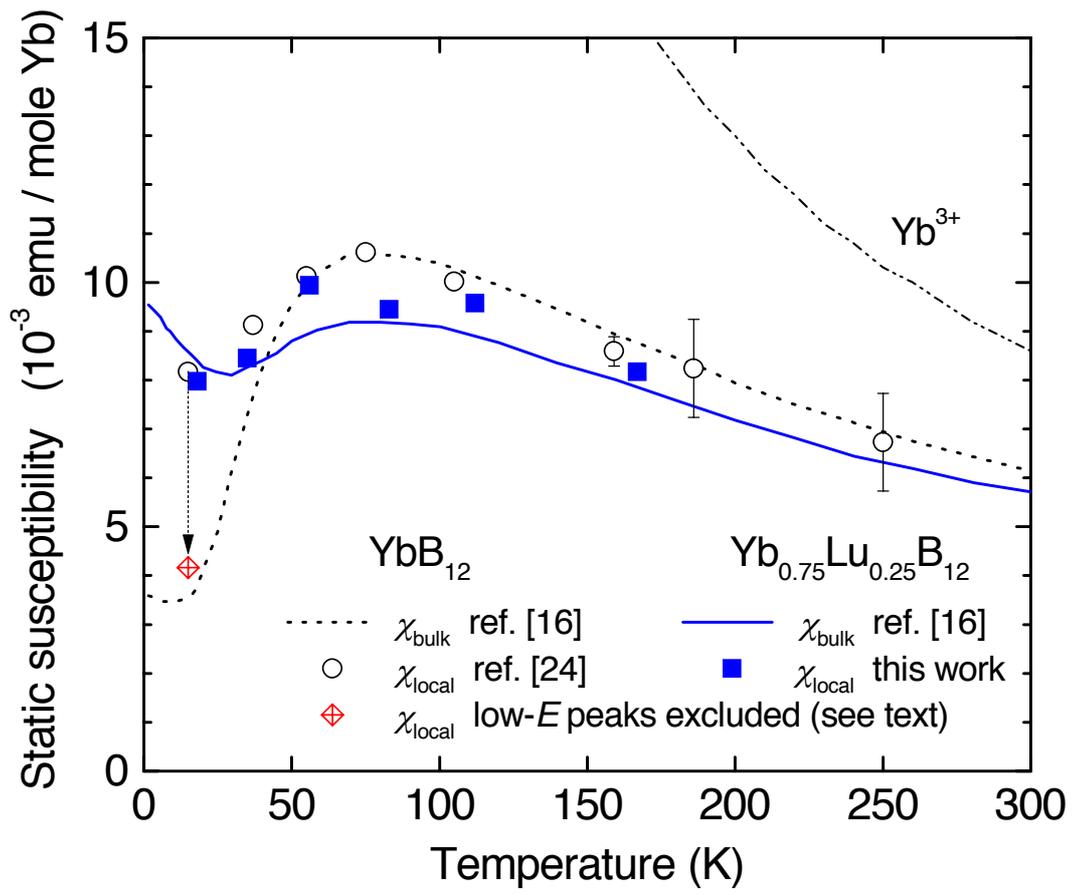

**Figure 9**